# Hybrid Parallel Quantum Key Distribution Protocol

Fábio Alencar Mendonça and Rubens Viana Ramos
alencar@deti.ufc.br          rubens@deti.ufc.br

Department of Teleinformatic Engineering, Faculty of Teleinformatic Engineering, Federal University of Ceará, 60455-760, C.P. 6007, Fortaleza-Ce, Brazil

**Abstract**

The practical realizations of BB84 quantum key distribution protocol using single-photon or weak coherent states have normally presented low efficiency, in the meaning that most bits sent by Alice are not useful for the final key. In this work, we show an optical setup to improve the transmission rate of useful bits putting together two ideas, parallel quantum key distribution and physical encryption using mesoscopic coherent states. The final result is a four time faster quantum key distribution setup.

## 1. Introduction

Quantum key distribution (QKD) is a technique based on quantum mechanics laws that permits two users of an authenticated channel to share, in a secure way, a random bit sequence used as key in cryptography algorithms. Due to the importance of confidentiality of the information trafficking in optical networks, QKD setups were rapidly implemented using light polarization [1-4], single-photon interference [5-9] and entangled pair of photons [10-12]. Currently, the single-photon interference setups are the most used, although the setups using polarization of single-photons have been proposed for fast (GHz clocked) and short distance QKD [13,14]. However, all the implemented QKD setups have presented low efficiency, in the meaning that most of the bits sent by the transmitter will not be used in the final key. This happens due to physical and logical reasons. The physical reasons are the losses in the optical fiber and optical devices, large number of empty pulses (vacuum states) when weak coherent states are used and low efficiency of single-photon detectors. The logical reason is the fact that QKD protocols require that only the bits in the time slots that transmitter and receiver have chosen the same bases be kept to form the key. Hence, at least 50% of the bits sent by the transmitter are lost, being the lower bond of 50% reached only for an ideal QKD setup.

Between the optical setups for QKD already proposed, a very interesting is one that uses single-photon interference in sidebands of phase or amplitude modulated light [15-19]. Recently it was shown that such system has a generalization that could, at least in principle, to permit to run two or more QKD protocols in parallel [20]. If two QKD protocols are run in parallel, then the transmission rate of useful bits can be doubled.

On the other hand, aiming to escape of the slow optical schemes based on single-photon sources and detectors, fast quantum communication setups using mesoscopic coherent states (QCMC) have been proposed [21-24]. Such system is based on the polarization of coherent states and it permits to reach gigabits rates in long distance optical networks. The security of the QCMC has been discussed in the literature [25-30] and, from our point of view based on the arguments used in favor and contrary, the system is secure for direct physical encryption but it is not secure for quantum key generation due to the use of a pseudo-random number generator. However, in order to circumvent this problem, recently it was proposed to use physical encryption and single-photon-based QKD together [31]. We call this system hybrid QKD because it uses weak (or single-photon) and mesoscopic coherent states. In this case, the data, randomly generated by Alice, are encrypted in the polarization of mesoscopic coherent states and are used to guide Bob in the choice of the bases used by him in a common QKD setup using weak coherent states. In this way, Alice and Bob choose always the same bases and none bit is lost in the sifting stage. This obviously doubles the useful bit transmission rate.

In this work we present a protocol for hybrid parallel QKD. In other words, we show how to implement two QKD protocols in parallel using single-photon interference in sidebands and physical encryption with mesoscopic coherent states. Such system has the transmission rate of useful bits four times larger than traditional QKD setup using single-photon interference in sidebands. This work is outlined as follows: in Section 2, the parallel QKD is proposed; In Section 3, physical encryption and hybrid QKD protocol are explained; In Section 4 the protocol for hybrid parallel QKD is presented and, at last, the conclusion are in Section 5.

## 2. Parallel QKD Protocol

The QKD setups using single-photon interference in sidebands can be constructed having phase modulators in Alice and Bob (PM-PM), amplitude modulators in Alice and Bob (AM-AM) or still amplitude (phase) modulator in Alice and phase (amplitude) modulator in Bob (AM-PM/PM-AM) [32]. Differently of [20], our parallel QKD system is of AM/PM type. The amplitude modulator used by Alice is a Mach-Zehnder (MZ) modulator. The proposed system can be seen in Fig. 1.

The electric field at MZ$_A$ output is given by

$$E_A = \frac{E_0 e^{j\omega_0 t}}{2}\left\{1 + e^{j\Psi_1} e^{j\left[m_1 \cos(\Omega_1 t + \Phi_{1A}) + m_2 \cos(\Omega_2 t + \Phi_{2A})\right]}\right\}, \tag{1}$$

where $\Psi_1$ is due to the DC polarization of the MZ modulator, $m_{1(2)}$ and $\Omega_{1(2)}$ are, respectively, the modulation depth and the radiofrequency for channels 1 and 2. Moreover, $\Phi_{1A}$ and $\Phi_{2A}$ are the phase shifts of the radiofrequencies $\Omega_1$ and $\Omega_2$ and they carry Alice's information. Having $m_{1(2)} \ll 1$, (1) can be simplified to

$$E_A = \frac{E_0 e^{j\omega_0 t}}{2}\left\{1 + e^{j\Psi_1}\left[1 + jm_1 \cos(\Omega_1 t + \Phi_{1A})\right]\left[1 + jm_2 \cos(\Omega_2 t + \Phi_{2A})\right]\right\}. \tag{2}$$

The field intensity can be calculated from (2), giving

$$I_A = \frac{E_0^2}{2}\left\{1 + \cos\Psi_1 - m_1 \sin\Psi_1 \cos(\Omega_1 t + \Phi_{1A}) - m_2 \sin\Psi_1 \cos(\Omega_2 t + \Phi_{2A})\right\}. \tag{3}$$

Choosing $\Psi_1 = 3\pi/2$, and substituting this value in (2) or (3), the following electric field is obtained.

$$E_A = \frac{E_0}{\sqrt{2}} \sqrt{1 + m_1 \cos(\Omega_1 t + \Phi_{1A}) + m_2 \cos(\Omega_2 t + \Phi_{2A})} e^{j\omega_0 t} \approx$$

$$\approx \frac{E_0}{\sqrt{2}} e^{j\omega_0 t} \left\{ 1 + \frac{m_1}{2} \cos(\Omega_1 t + \Phi_{1A}) + \frac{m_2}{2} \cos(\Omega_2 t + \Phi_{2A}) \right\}. \qquad (4)$$

In (4) the global phase depending on $\Omega_{1,2}$ and $\Phi_{1A,2A}$ was not considered. After propagation in an optical fiber link with length $L$, the electric field at Bob's phase modulator input is

$$E_F = \frac{E_0}{\sqrt{2}} \left\{ e^{j(\beta_0 L + \omega_0 t)} + \frac{m_1}{4} \left[ e^{j\left(\beta_1^+ L + (\omega_0 + \Omega_1)t + \Phi_{1A}\right)} + e^{-j\left(-\beta_1^- L - (\omega_0 - \Omega_1)t + \Phi_{1A}\right)} \right] \right\}$$

$$+ \frac{E_0}{\sqrt{2}} \frac{m_2}{4} \left[ e^{j\left(\beta_2^+ L + (\omega_0 + \Omega_2)t + \Phi_{2A}\right)} + e^{-j\left(-\beta_2^- L - (\omega_0 - \Omega_2)t + \Phi_{2A}\right)} \right], \qquad (5)$$

where $\beta_0$, $\beta_{1(2)}^+$ e $\beta_{1(2)}^-$ are the propagation constants for each frequency component of $E_F$: $\omega_0$, $\omega_0 + \Omega_1$, $\omega_0 - \Omega_1$, $\omega_0 + \Omega_2$, $\omega_0 - \Omega_2$. Despising chromatic dispersion, $\beta_{1(2)}^\pm$ are given by

$$\beta_{1(2)}^\pm = \frac{n}{c}(\omega_0 \pm \Omega_{1(2)}) = \beta_0 \pm \frac{n}{c} \Omega_{1(2)}. \qquad (6)$$

Substituting (6) in (5) and taking off the global phase $e^{j\beta_0 L}$, (5) can rewritten as

$$E_F = \frac{E_0}{\sqrt{2}} \left\{ e^{j\omega_0 t} + \frac{m_1}{4} \left[ e^{j\left(\frac{n}{c}\Omega_1 L + (\omega_0 + \Omega_1)t + \Phi_{1A}\right)} + e^{-j\left(\frac{n}{c}\Omega_1 L - (\omega_0 - \Omega_1)t + \Phi_{1A}\right)} \right] + \frac{m_2}{4} \left[ e^{j\left(\frac{n}{c}\Omega_2 L + (\omega_0 + \Omega_2)t + \Phi_{2A}\right)} + e^{-j\left(\frac{n}{c}\Omega_2 L - (\omega_0 - \Omega_2)t + \Phi_{2A}\right)} \right] \right\}. \quad (7)$$

Now, after Bob's phase modulator the electric field is

$$E_B = E_F e^{j\left[m_3 \cos(\Omega_1 t + \Phi_{1B}) + m_4 \cos(\Omega_2 t + \Phi_{2B})\right]}. \quad (8)$$

In (8) $m_{3(4)}$ are, respectively, the modulation depths used by Bob for the channels 1 and 2. Moreover, $\Phi_{1B}$ and $\Phi_{2B}$ are the phases of the radiofrequencies $\Omega_1$ and $\Omega_2$, respectively, chosen by Bob according to QKD protocol. Considering $m_{3(4)} \ll 1$, (8) can be approximated by

$$\begin{aligned} E_B &= E_F \left[1 + jm_3 \cos(\Omega_1 t + \Phi_{1B})\right]\left[1 + jm_4 \cos(\Omega_2 t + \Phi_{2B})\right] \\ &= E_F \left[ \begin{array}{l} 1 + jm_3 \cos(\Omega_1 t + \Phi_{1B}) + jm_4 \cos(\Omega_2 t + \Phi_{2B}) - \\ - m_3 m_4 \cos(\Omega_1 t + \Phi_{1B})\cos(\Omega_2 t + \Phi_{2B}) \end{array} \right]. \end{aligned} \quad (9)$$

Substituting (7) in (9) one finds

$$E_F = \frac{E_0}{\sqrt{2}} \left\{ \begin{array}{l} e^{j\omega_0 t} + \frac{m_1}{4}\left[ e^{j\left(\frac{n}{c}\Omega_1 L + (\omega_0+\Omega_1)t + \Phi_{1A}\right)} + e^{-j\left(\frac{n}{c}\Omega_1 L - (\omega_0-\Omega_1)t + \Phi_{1A}\right)} \right] + \\ + \frac{m_2}{4}\left[ e^{j\left(\frac{n}{c}\Omega_2 L + (\omega_0+\Omega_2)t + \Phi_{2A}\right)} + e^{-j\left(\frac{n}{c}\Omega_2 L - (\omega_0-\Omega_2)t + \Phi_{2A}\right)} \right] \end{array} \right\} \times$$

$$\left\{ \begin{array}{l} 1 + jm_3 \cos(\Omega_1 t + \Phi_{1B}) + jm_4 \cos(\Omega_2 t + \Phi_{2B}) - \\ - m_3 m_4 \cos(\Omega_1 t + \Phi_{1B})\cos(\Omega_2 t + \Phi_{2B}) \end{array} \right\}. \qquad (10)$$

Finally, the intensities of the sidebands for each modulating frequency $\Omega_{1(2)}$ can be calculated from (10). The results are

$$I_B^{(\omega_0 \pm \Omega_1)} = \frac{E_0^2}{8}\left[ \frac{m_1^2}{4} + m_3^2 \pm m_1 m_3 \sin\left(\frac{n}{c}\Omega_1 L + \Delta\Phi_1\right) \right] \qquad (11)$$

$$I_B^{(\omega_0 \pm \Omega_2)} = \frac{E_0^2}{8}\left[ \frac{m_2^2}{4} + m_4^2 \pm m_2 m_4 \sin\left(\frac{n}{c}\Omega_2 L + \Delta\Phi_2\right) \right], \qquad (12)$$

where $\Delta\Phi_1 = \Phi_{1A} - \Phi_{1B}$ and $\Delta\Phi_2 = \Phi_{2A} - \Phi_{2B}$. Observing (11) and (12), one can see that interference occurs independently for each channel. Furthermore, if $(n/c)\Omega_1 L = \pi/2$, $(n/c)\Omega_2 L = 3\pi/2$, $m_1 = 2m_3$ and $m_2 = 2m_4$, (11) and (12) are of the form

$$I_B^{(\omega_0 + \Omega_1)} = \frac{E_0^2 m_1^2}{16} \cos^2\left(\frac{\Phi_{1A} - \Phi_{1B}}{2}\right) \qquad (13)$$

$$I_B^{(\omega_0 - \Omega_1)} = \frac{E_0^2 m_1^2}{16} \operatorname{sen}^2\left(\frac{\Phi_{1A} - \Phi_{1B}}{2}\right) \qquad (14)$$

$$I_B^{(\omega_0+\Omega_2)} = \frac{E_0^2 m_2^2}{16}\operatorname{sen}^2\left(\frac{\Phi_{2A}-\Phi_{2B}}{2}\right) \tag{15}$$

$$I_B^{(\omega_0-\Omega_2)} = \frac{E_0^2 m_2^2}{16}\cos^2\left(\frac{\Phi_{2A}-\Phi_{2B}}{2}\right) \tag{16}$$

Hence, using the system of Fig. 1, one can run two BB84 protocols in parallel, one in channel 1 and the other in channel 2, doubling the transmission rate of useful bits.

## 3. Physical Encryption and Hybrid QKD Protocol

The unconditional security of QCMC is still an opened question [25-30]. From our point of view, QCMC is not secure for QKD but it is secure for direct physical encryption. This mean that Alice can send a secret message to Bob and an eavesdropper will not be able to read it correctly, but such message cannot be used as key in a cryptography algorithm. In order to go a little deeper in the security analysis, let us consider the brute-force attack in the QCMC using linear polarization of two-mode coherent states. Firstly, any two-mode coherent state having linear polarization can constructed applying a polarization rotation in the horizontal state $|\alpha,0\rangle$

$$R(\theta)|\alpha,0\rangle = \exp\left(-\theta\left(\hat{a}_V^+\hat{a}_H - \hat{a}_H^+\hat{a}_V\right)\right)|\alpha,0\rangle = |\alpha\cos(\theta),\alpha\sin(\theta)\rangle, \tag{17}$$

where $\hat{a}_V$ and $\hat{a}_H$ are, respectively, the annihilation operators of the vertical and horizontal modes. For this linearly polarised light, the mean values and the variances of the Stokes parameters are known to be $\langle S_1\rangle=|\alpha|^2\cos(2\theta)$, $\langle S_2\rangle=|\alpha|^2\sin(2\theta)$, $\langle S_3\rangle=0$ and

$V_{S_1} = V_{S_2} = V_{S_3} = |\alpha|^2$. Hence, the uncertainty in the polarization depends directly on the intensity of the optical field. A related question is how good one can distinguish between two linear polarisation states having a dephasing of $\theta$ between them. This measure is given by the inner product and, without loss of generality, let us consider one of the polarisations to be the linear horizontal state. The inner product is then given by:

$$D = \left|\langle \alpha, 0 | R(\theta) | \alpha, 0 \rangle\right|^2 = \exp\left(-2|\alpha|^2 \sin^2(\theta)\right). \tag{18}$$

Hence, the smaller (toward zero) the angle $\theta$ the larger the necessary mean photon number of the coherent state for a good distinguishability. This point is crucial for the security of the quantum communication protocols based on mesoscopic coherent states, since it states that some relation between the mean photon number and the number of bases used must be obeyed in order to guarantee the security of the protocol. This fact can be well understood analysing the brute force attack. Let us suppose that Alice chooses randomly (with equal probability) a linearly polarized two-mode coherent state from the set $\{|\theta_0\rangle,...,|\theta_{2M-1}\rangle\}$, $0 \leq \theta \leq \pi$, hence, there are $M$ bases. The goal of Eve is to identify the polarization sent by Alice. If the pulse has too many photons, Eve can apply the brute force attack, that is, she divide the pulse sent by Alice in $M$ other pulses with the same polarization. For the $P_i$ ($i=0...M-1$) pulse, Eve applies the polarization rotation of -$\phi_i$, $0 \leq \phi \leq \pi/2$, and passes it through a polarization beam splitter (PBS) having single-photon counters at both PBS outputs. The optical setup is shown in Fig. 2.

If Eve has perfect detectors (unitary efficiency and noiseless), the following cases are possible:

1) Detection in both detectors: In this case Eve knows the polarization rotation is wrong.

2) None detection in both detectors: In this case Eve does not gain any information.

3) Detection in only one detector: In this case there are two possibilities; Eve used the correct polarization rotation or Eve used a wrong polarization rotation but there was no photon in one of the PBS output.

Hence, if the optical pulse has a number of photons much larger than $M$, with high probability Eve will get the right polarization. On the other hand, if the number of photons of the pulse is lower than $M$, then Eve will not have enough photons to test all possible polarization rotations and she will not be able to determine, with good precision, the linear polarization sent by Eve. Hence, in order to implement a secure quantum communication system where the bits are coded in the polarization the condition $|\alpha|^2 \ll M$ must be satisfied. Now, one can argue that Eve can use an optical amplifier in order to have $|\alpha|^2 \gg M$. However, an optical amplifier will, unavoidably, to produce unpolarized photons due to generation and amplification of spontaneous emissions (ASE). These upolarized photons will cause mistakes in Eve's detection and, once in Bob, they will cause counts in the detector that he was not expecting to receive any photon, since in the QCMC Bob knows in advance the polarizations used by Alice.

The hybrid QKD protocol is described as follow:

1. Alice and Bob share in advance a secret key $K$ having $|K|$ bits. Using this key as seed in a pseudo-random number generator, a new expanded key $K'$ having $|K'|$ bits is created.

2. Alice generates a true random sequence of bits $R$ having $|R| = \lfloor |K'|/\log_2(M) \rfloor$ ($\lfloor o \rfloor$ means the first integer lower than 'o') bits. Thus, for each bit of $R$, Alice uses $\log_2(M)$ bits of $K'$, that has a decimal representation $D_K$, to choose the basis of polarization to be used. Further, if $D_K$ is an even number and the bit value of $R$ is '0', then the polarization in the first quadrant is used ($<\pi/2$). If bit value of $R$ is '1', then the polarization in the second quadrant is used ($>\pi/2$). On the other hand, if $D_K$ is an odd number and the bit value of $R$ is '0', then the polarization in the second quadrant is used ($>\pi/2$). If bit value of $R$ is '1', then the polarization in the first quadrant is used ($<\pi/2$). Obeying the just described codification, Alice sends linearly polarized optical pulses to Bob. The mean photon number of the pulses sent by Alice are much lower than $M$.

3. Since Bob knows the expanded key $K'$, he will always measure the optical pulse sent by Alice in the correct polarization basis and he will know which polarizations mean bit '0' and '1'. Hence, Bob will obtain the correct bit sequence $R$ sent by Alice with high probability.

4. The BB84 protocol is realized with Alice and Bob using the bits of $R$ to choose the bases.

There are two clear advantages in this hybrid QKD protocol. Firstly, the sifting stage does not exist anymore, since Alice and Bob always choose the same bases. Obviously, this doubles the transmission rate of useful bits. Secondly, at the photon number splitting attack (PNS) [33,34], Eve keeps one photon and sends the rest of the pulse to Bob. She maintains her photon in a quantum memory waiting Alice and Bob divulgate publicly the bases used. However, using the hybrid QKD protocol, Alice and Bob will not divulgate the basis used anymore. Hence, quantum memory will be useless for Alice and the PNS will be harder because Eve will need pulses having at least three photons. Then, the best that Eve can do is to attack the mesoscopic coherent states but, as explained above and in the references [25,26,28,31], she will not have success. Moreover, Bob can monitor the optical power of the mesoscopic pulses and not expected detections in his detectors in order to infer if Eve applied an attack using a beam-splitter or beam splitter/optical amplifier attack. Obviously, this is possible if there are not optical amplifiers in the link between Alice and Bob, otherwise Bob will not know if the unpolarized photons are due to Eve's attack or in line amplifiers. An example of hybrid QKD implementation can be seen in Fig. 3. It is a polarimetric BB84 protocol.

In Fig. 3, EOS is an electro-optic switch and HWP is a half wave plate. As can be seen in Fig. 3, the 1300 nm channel is used for QCMC employing passive linear optical error correction [35,36], while the 850 nm channel is used for fast and short distance polarimetric QKD.

## 4. Hybrid Parallel QKD Protocol

The construction of a hybrid parallel QKD protocol is, now, straightforward. Alice and Bob have a QKD setup as shown in Fig. 1 and a QCMC setup as shown in the 1300 nm channel of Fig. 3. The protocol is exactly as explained in steps 1-4 of Section 3. In the fourth step, Alice and Bob use bit sequence $R$ to choose the bases for both QKD systems, channel 1 (radiofrequency $\Omega_1$) and channel 2 (radiofrequency $\Omega_2$). Since each QKD system will double its transmission rate of useful bits and there are two QKD systems working in parallel, if the final key is composed by concatenation of both individual keys, the total transmission rate of useful bits will be four time larger than a traditional BB84 QKD. The optical setup for hybrid parallel QKD can be seen in Fig. 4.

## 5. Conclusions

Firstly, we have shown that it is possible to implement parallel QKD using single-photon interference in sidebands with two carriers, employing amplitude modulator in Alice and phase modulator in Bob. Following, we showed how to construct a hybrid QKD protocol. The security against brute force attack was discussed and an optical setup was presented. At last, we put everything together, showing how to implement hybrid parallel QKD having transmission rate of useful bits four times larger than traditional weak coherent state-based QKD.

## Acknowledgments

This work was supported by the Brazilian agency FUNCAP.

# Figure 1

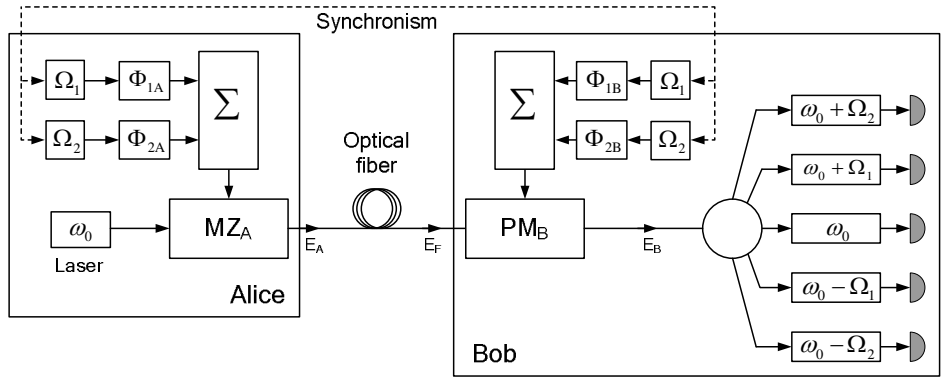

Figure 1: Optical setup for parallel QKD using two carriers.

**Figure 2**

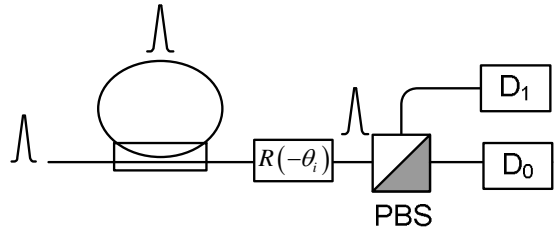

Figure 2: Linear polarization identification using brute force.

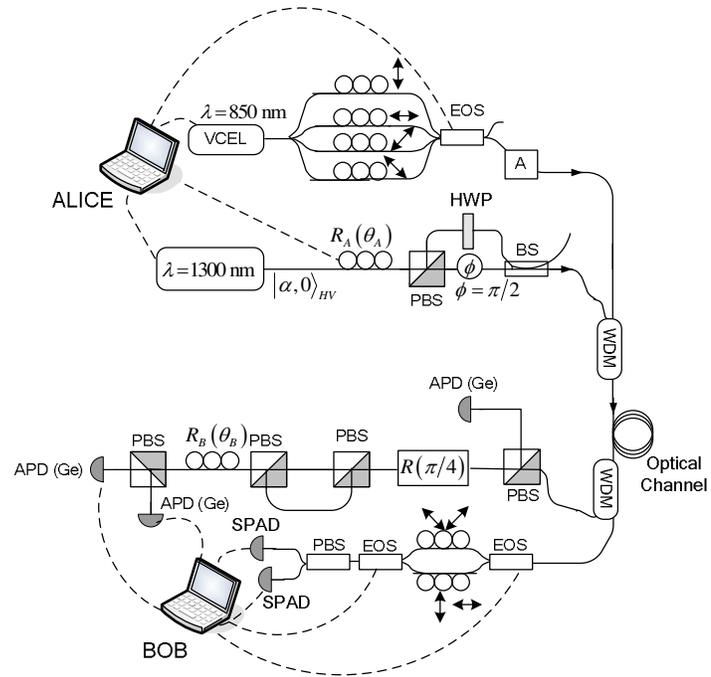

Figure 3: Polarimetric hybrid QKD optical setup.

**Figure 4**

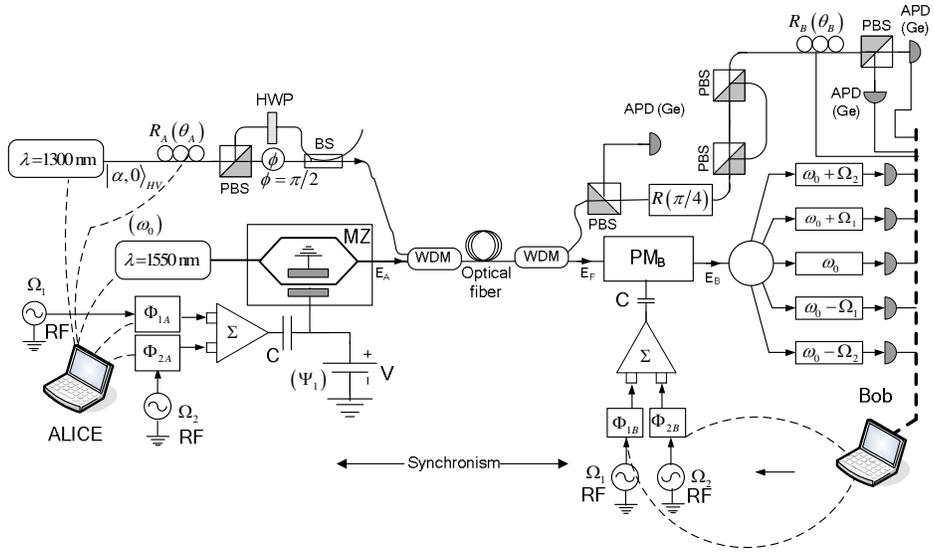

Figure 4: Optical setup for hybrid parallel QKD.